\documentclass[a4paper,11pt]{article}
\pdfoutput=1 
\usepackage{jheppub} 
\usepackage{bbm,bm,graphicx,mathtools,color,hyperref,slashed}
\usepackage{amssymb,amsmath}
\usepackage[T1]{fontenc}

\usepackage[all]{xy}

\graphicspath{{./Figures/}}

\newcommand{\diff}{\mathrm{d}}
\newcommand{\p}{\partial}

\newcommand{\Diff}{{\mathcal{D}}}

\newcommand{\be}{\begin{equation}}      
\newcommand{\ee}{\end{equation}}      
\newcommand{\bea}{\begin{eqnarray}}      
\newcommand{\eea}{\end{eqnarray}}

\newcommand{\tr}{\mathrm{tr}}

\newcommand{\im}{\mathrm{i}}

\newcommand{\calA}{\mathcal{A}}
\newcommand{\calB}{\mathcal{B}}

\newcommand{\calL}{\mathcal{L}}
\newcommand{\calR}{\mathcal{R}}

\newcommand{\calM}{\mathcal{M}}

\newcommand{\rmc}{\mathrm{c}}
\newcommand{\rme}{\mathrm{e}}
\newcommand{\rmf}{\mathrm{f}}
\newcommand{\rmA}{\mathrm{A}}
\newcommand{\rmB}{\mathrm{B}}
\newcommand{\rmL}{\mathrm{L}}
\newcommand{\rmR}{\mathrm{R}}
\newcommand{\rmV}{\mathrm{V}}

\title{Anomaly constraint on massless QCD and the role of Skyrmions in chiral symmetry breaking}

\author{Yuya Tanizaki}

\affiliation{RIKEN BNL Research Center, Brookhaven National Laboratory, Upton, NY 11973 USA}

\emailAdd{yuya.tanizaki@riken.jp}

\abstract{
We discuss consequences of the 't Hooft anomaly matching condition for Quantum Chromodynamics (QCD) with massless fundamental quarks. 
We derive the new discrete 't Hooft anomaly of massless QCD for generic numbers of color $N_\mathrm{c}$ and flavor $N_\mathrm{f}$, and an exotic chiral-symmetry broken phase without quark-bilinear condensate is ruled out from possible QCD vacua.  
We show that the $U(1)_\mathrm{B}$ baryon number symmetry is anomalously broken when the $(\mathbb{Z}_{2N_\mathrm{f}})_\mathrm{A}$ discrete axial symmetry and the flavor symmetry are gauged. 
In the ordinary chiral symmetry breaking, the Skyrmion current turns out to reproduce this 't Hooft anomaly of massless QCD. 
In the exotic chiral symmetry breaking, however, the anomalous breaking of $U(1)_\mathrm{B}$ does not take the correct form, and it is inconsistent with anomaly matching. This no-go theorem is based only on symmetries and anomalies, and thus has a wider range of applicability to the QCD phase diagram than the previous one obtained by QCD inequalities. 
Lastly, as another application, we check that duality of $\mathcal{N}=1$ supersymmetric QCD with $N_\mathrm{f}\ge N_\mathrm{c}+1$ satisfies the new anomaly matching. 
}

\begin{document}
\maketitle
\section{Introduction}\label{sec:introduction}

Quantum chromodynamics (QCD) is a part of the Standard Model of particle physics that describes the fundamental law of quarks and gluons. 
It is the vector-like $SU(N_\rmc)$ gauge theory coupled to Dirac fermions in the defining representation $\bm{N_\rmc}$, and provides the complete foundation for strong interaction of nuclear physics. 
Therefore, solving QCD is the ultimate goal of nuclear and hadron physics, but it is quite a difficult task because the theory becomes strongly coupled at low-energies $E\lesssim \Lambda_{\mathrm{QCD}}$ due to the asymptotic freedom. 

In order to understand properties of QCD, symmetry has  been playing a pivotal role. In our universe, the up and down (also strange) quarks are light compared with $\Lambda_{\mathrm{QCD}}$, and the theory has an approximate chiral symmetry $SU(N_\rmf)_\rmL\times SU(N_\rmf)_\rmR$ with $N_\rmf=2$ (or $3$). 
Nambu and Jona-Lasinio showed in a model of four-fermion interaction that if the interaction between quarks are sufficiently strong then this chiral symmetry is spontaneously broken to the diagonal subgroup~\cite{Nambu:1961tp, Nambu:1961fr},  
\be
SU(N_\rmf)_\rmL\times SU(N_\rmf)_\rmR\to SU(N_\rmf)_\rmV. 
\ee
Assuming the symmetry breaking pattern, the method of phenomenological Lagrangian tells us that the low-energy behavior of the (pseudo) Nambu-Goldstone bosons is determined without knowing details of dynamics~\cite{Weinberg:1966fm, Schwinger:1967tc, Coleman:1969sm, Callan:1969sn}, and it has been confirmed that light pseudo-scalar mesons $\pi$ (and $K,\eta$) are Nambu-Goldstone bosons of chiral symmetry breaking. 
This is a great success of hadron physics by paying attention to chiral symmetry of QCD, while it remains an open question why QCD vacuum breaks chiral symmetry.

The question is partly solved by 't~Hooft anomaly matching argument~\cite{tHooft:1979rat, Frishman:1980dq, Coleman:1982yg}, so let us review it using modern terminologies~\cite{Kapustin:2014lwa, Kapustin:2014zva}. 
We consider QCD with $N_\rmf$ massless quarks, then it has chiral symmetry $SU(N_\rmf)_\rmL\times SU(N_\rmf)_\rmR$ and this symmetry is well-defined as global symmetry, i.e. not broken by quantum anomaly or QCD instantons. 
We put the theory on a closed four-manifold $M_4$ and introduce the background $SU(N_\rmf)_{\rmL,\rmR}$ gauge fields, $L, R$, to define the partition function $Z_{M_4}[L,R]$. 
However, $Z_{M_4}[L,R]$ breaks the gauge invariance in terms of $L$ and $R$, and this violation of the background-field gauge invariance is recently called 't~Hooft anomaly. 
The descent procedure on Stora-Zumino chain~\cite{Stora:1976kd, Stora:1983ct, Zumino:1983ew} says that gauge-invariance for $L, R$ is established by adding $5$-dimensional Chern-Simons action $\mathrm{CS}_{M_5}[L,R]$ of the background gauge field $L,R$. Therefore, we have to introduce the auxiliary $5$-dimensional spacetime $M_5$ with $\p M_5=M_4$, and the combined system
\be
Z_{M_4}[L,R]\exp\left[\im N_c \left(\mathrm{CS}_{M_5}[L]-\mathrm{CS}_{M_5}[R]\right)\right], 
\ee
is gauge invariant.
Making our spacetime $M_4$ sufficiently large, the partition function $Z_{M_4}[L,R]$ is effectively described only by massless states of the physical Hilbert space. The anomaly inflow~\cite{Callan:1984sa} from the auxiliary $5$-dimensional bulk requires that the low-energy effective theory of massless particles reproduce the same 't~Hooft anomaly. 
That is, 't~Hooft anomaly is invariant under the renormalization-group flow, and this is the 't~Hooft anomaly matching condition. 
In order to match the anomaly in the infrared limit, the unique and gapped ground state is ruled out from possible vacuum structures, and we conclude 
\begin{itemize}
\item the low-energy effective theory is conformal, or 
\item chiral symmetry is spontaneously broken and massless pions exist. 
\end{itemize}
In the second scenario, the low-energy Lagrangian of pions must contain the Wess-Zumino term~\cite{Wess:1971yu, Witten:1983tw} to match the 't~Hooft anomaly, and this is crucial for correct description of neutral pion decays, which historically determines $N_\rmc=3$~\cite{Adler:1969gk, Bell:1969ts}. 
The above argument elucidates that chiral symmetry breaking is partly required by the topological nature of chiral symmetry when massless fermions exist in the ultraviolet theory, i.e. QCD Lagrangian. 

In this paper, we shall derive a new 't~Hooft anomaly by looking more carefully at the symmetry of massless QCD, and put a stronger constraint on the possible low-energy dynamics. 
This becomes possible thanks to the recent development about understanding of 't~Hooft anomaly matching as a nontrivial surface state of symmetry-protected topological (SPT) orders~\cite{Vishwanath:2012tq, Wen:2013oza, Cho:2014jfa, Kapustin:2014lwa, Kapustin:2014zva, Wang:2014pma}, which elucidates the deep connection between 't~Hooft anomaly matching~\cite{tHooft:1979rat, Frishman:1980dq, Coleman:1982yg} in high-energy physics and the Lieb-Schultz-Mattis theorem~\cite{Lieb:1961fr, Affleck:1986pq, PhysRevLett.84.1535, Hastings:2003zx} in condensed matter physics.
This enables us to apply 't~Hooft anomaly matching condition for various symmetries, including discrete symmetries, symmetries with projective realizations, higher-form symmetries, and so on, and many nonperturbative aspects of quantum field theories are newly discovered~\cite{Csaki:1997aw, Witten:2015aba,Seiberg:2016rsg,Witten:2016cio, Tachikawa:2016cha, Tachikawa:2016nmo, Gaiotto:2017yup,Wang:2017txt, Tanizaki:2017bam, Komargodski:2017dmc, Komargodski:2017smk, Cho:2017fgz,Shimizu:2017asf, Wang:2017loc,Kikuchi:2017pcp, Gaiotto:2017tne, Gomis:2017ixy, Tanizaki:2017qhf, Tanizaki:2017mtm, Cherman:2017dwt, Yamazaki:2017dra, Guo:2017xex, Dunne:2018hog, Sulejmanpasic:2018upi, Cordova:2018cvg, Aitken:2018kky, Kobayashi:2018yuk, Tanizaki:2018xto, Cordova:2018acb, Anber:2018iof, Anber:2018jdf}.

For our purpose, it is important to identify the internal symmetry with the faithful representation  correctly, and the symmetry of massless QCD is given by the quotient group, 
\bea
G&=&{SU(N_\rmf)_\rmL\times SU(N_\rmf)_\rmR\times U(1)_\rmV\times (\mathbb{Z}_{2 N_\rmf})_\rmA\over \mathbb{Z}_{N_\rmc}\times (\mathbb{Z}_{N_\rmf})_\rmL\times (\mathbb{Z}_{N_\rmf})_\rmR\times \mathbb{Z}_2}\\
&=&{SU(N_\rmf)_\rmL\times SU(N_\rmf)_\rmR\times U(1)_\rmV\over \mathbb{Z}_{N_\rmc}\times (\mathbb{Z}_{N_\rmf})_{\rmV}}, 
\eea
which meaning shall be explained in Sec.~\ref{sec:symmetry_massless_QCD}. 
When properly gauging the quotient group $G$, we have to introduce not only ordinary one-form gauge fields but also the two-form gauge fields, and both of them become equally important.
The original 't~Hooft anomaly matching is only sensitive to the infinitesimal part of $G$ around identity, and we show that non-trivial topology of $G$ provides a new anomaly matching condition. In this paper, we especially consider the subgroup, 
\be
G^{\mathrm{sub}}={SU(N_\rmf)_\rmV\times U(1)_\rmV\over \mathbb{Z}_{N_\rmc}\times (\mathbb{Z}_{N_\rmf})_\rmV} \times (\mathbb{Z}_{N_\rmf})_\rmL \subset G, 
\ee
and discuss the consequence of the 't~Hooft anomaly of $G^{\mathrm{sub}}$. 

The main outcome of our computation in Sec.~\ref{sec:anomaly} is that massless QCD has a new 't~Hooft anomaly, characterized by $5$-dimensional topological action,
\be
S_{\mathrm{SPT}}={N_\rmf\over (2\pi)^2}\int_{M_5} A^{(1)}_\chi\wedge \diff A_\rmB \wedge B^{(2)}_\rmf+\cdots. 
\ee
Here, $A^{(1)}_\chi$ is the $(\mathbb{Z}_{N_\rmf})_\rmL$ gauge field, $A_\rmB$ is the $U(1)_\rmB=U(1)_\rmV/\mathbb{Z}_{N_\rmc}$ gauge field, and $B^{(2)}_\rmf$ is the $(\mathbb{Z}_{N_\rmf})_{\rmV}$ two-form gauge field. 
This is the mixed 't~Hooft anomaly involving the discrete chiral symmetry $(\mathbb{Z}_{N_\rmf})_\rmL$, the baryon-number symmetry $U(1)_\rmB$, and the projective nature of the vector-like flavor symmetry $[SU(N_\rmf)_\rmV\times U(1)_\rmB]/(\mathbb{Z}_{N_\rmf})_\rmV$. 
The anomaly matching condition claims that the low-energy effective theory of massless QCD must reproduces the same anomaly. 
The related anomaly is partly discussed in some recent studies~\cite{Tanizaki:2017bam, Shimizu:2017asf,  Gaiotto:2017tne, Tanizaki:2017qhf, Tanizaki:2017mtm, Cherman:2017dwt}, but all of them require nontrivial common divisor of $N_\rmc$ and $N_\rmf$, $\mathrm{gcd}(N_\rmc,N_\rmf)>1$, such as $N_\rmc=N_\rmf$. In this paper, we remove this constraint, which is important since our universe is closest to the case $N_\rmc=3$ and $N_\rmf=2$, and also clarify the physical meaning. 

In Sec.~\ref{sec:anomaly_matching}, we first consider the ordinary scenario of chiral symmetry breaking by the quark bilinear condensate $\langle \overline{\psi}\psi\rangle \not=0$, and discuss how the anomaly matching can be satisfied. 
We can check that the most term of the anomaly is matched by the Wess-Zumino term, but the anomaly of $S_{\mathrm{SPT}}$ cannot be matched only by it because pions do not have the baryon charge but $S_{\mathrm{SPT}}$ requires the nontrivial action under $U(1)_\rmB$. 
We show that the solution is given by the unified description of nucleons and pions proposed by Skyrme~\cite{Skyrme:1961vq, Skyrme:1962vh}. Indeed, Skyrme noticed, even before the establishment of QCD, that the nonlinear sigma model unifies the theory of mesons and baryons, where baryons are given by topologically stable configuration of the nonlinear field. 
Since chiral symmetry is broken as 
\be
G\to H={SU(N_\rmf)_\rmV\times U(1)_\rmV\over \mathbb{Z}_{N_\rmc}\times (\mathbb{Z}_{N_\rmf})_\rmV}, 
\ee
the target space of the nonlinear sigma model is $G/H\simeq SU(N_\rmf)$. The topologically stable soliton is characterized by $\pi_3(G/H)=\mathbb{Z}$, and this integer is nothing but the baryon number of $U(1)_\rmB$. 
Heuristically, this can be understood that the theory is still described only by pions when taking the low-energy limit under non-zero baryon numbers and the baryons can be seen as the topological defect of the pion fields. 
Since our anomaly matching condition requires the nontrivial action of $U(1)_\rmB$ to the Hilbert space of low-energy effective theory, this heuristic argument is promoted to the rigorous consequence for the chiral-symmetry broken phase of massless QCD: Nontrivial homotopy, $\pi_3$, of the vacuum manifold is designated by anomaly matching.
Here, let us remind that the corresponding Noether current $J_\rmB$ is given by three-dimensional Wess-Zumino term, and its minimal coupling to $A_\rmB$ is important to reproduce the $U(1)_\rmB$-$SU(N_\rmf)_\rmL$-$SU(N_\rmf)_\rmL$ triangle anomaly, as pointed out by Witten~\cite{Witten:1983tw}. 
Our anomaly further requires that $J_\rmB$ must correctly transform under the discrete chiral symmetry, and we show that this is indeed the case for the ordinary chiral broken phase. 

In Sec.~\ref{sec:anomaly_matching}, we also examine and rule out an exotic scenario of chiral symmetry breaking, originally proposed by Stern~\cite{Stern:1997ri, Stern:1998dy, Kogan:1998zc, Kanazawa:2015kca}. This can be characterized by the symmetry breaking pattern,
\be
G\to G^{\mathrm{sub}}, 
\ee
so the target space is given by $G/G^{\mathrm{sub}}\simeq PSU(N_\rmf)=SU(N_\rmf)/\mathbb{Z}_{N_\rmf}$.  
The most striking difference from the ordinary scenario is that the vacuum is invariant under the discrete chiral symmetry $(\mathbb{Z}_{N_\rmf})_\rmL$, and thus quark bilinear condensate must disappear, $\langle \overline{\psi}\psi\rangle=0$. 
This exotic phase has been ruled out by QCD inequality~\cite{Kogan:1998zc}, but it is still an open problem whether it appears at finite-density massless QCD, because the QCD inequality cannot be applied with the sign problem. 
In this work, we rule out this exotic phase from possible zero-temperature QCD vacua  since the Noether current $J_\rmB$ of $U(1)_\rmB$ does not obey the required transformation law under the discrete axial symmetry $(\mathbb{Z}_{N_\rmf})_\rmL$. 
As the argument relies only on symmetry and anomaly, our no-go theorem applies to much wider region of the QCD phase diagram, especially the zero-temperature finite-density QCD. 

In Sec.~\ref{sec:seiberg_duality}, we consider the anomaly matching condition for $\mathcal{N}=1$ supersymmetric QCD with $N_\rmf\ge N_\rmc+1$. When $N_\rmf\ge N_\rmc+2$, $SU(N_\rmc)$ SQCD is mapped to $SU(N_\rmf-N_\rmc)$ SQCD by Seiberg duality,  and we can explicitly check that these theories have the same new 't~Hooft anomaly. 
We also consider the $s$-confining phase at $N_\rmf=N_\rmc+1$, and we see that the massless baryons correctly satisfy the new anomaly matching. 
These examples give a good lesson about how the new anomaly matching condition can be satisfied in chiral symmetric phases. 

\section{Symmetry of massless QCD}\label{sec:symmetry_massless_QCD}
We consider the four-dimensional gauge theory with the gauge group $SU(N_\rmc)$ coupled to $N_\rmf$ massless Dirac fermions in the fundamental representation ($N_\rmf$-flavor massless QCD). 
The classical action of this theory is given by 
\be
S={1\over 2g^2}\int \tr_\rmc(F_\rmc(a)\wedge \star F_\rmc(a))+\int \overline{\psi}\gamma_{\mu}D_{\mu}(a)\psi. 
\label{eq:classical_action_QCD}
\ee
Here, $a$ is the $SU(N_\rmc)$ gauge field ($a^\dagger =-a$)\footnote{Throughout this paper, we follow the convention that the dynamical gauge fields are denoted by lowercases $a,b,\ldots$, and the background gauge fields are by uppercases, $A,B,\ldots$. The gauge fields in our convention are realized locally as anti-Hermitian matrix-valued one-forms. }, $D(a)=\diff+a$ is the covariant derivative, $F_\rmc(a)= D(a)\wedge D(a)=\diff a+a \wedge a$ is the $SU(N_\rmc)$ gauge field strength, $\psi$ is the quark field realized as $N_\rmc\times N_\rmf$ Grassmannian variables, $\overline{\psi}$ is the conjugate field of $\psi$, and $\tr_{\rmc}$ represents the trace over color indices in the defining representation. 
When it is evident, we simply write $D=D(a)$ and $F_\rmc=F_\rmc(a)$.  

\subsection{Symmetry group of massless $N_\rmf$-flavor QCD}

The (internal) global symmetry of this theory is given by 
\be
G=\frac{SU(N_\rmf)_\rmL \times SU(N_\rmf)_\rmR\times U(1)_\rmV\times (\mathbb{Z}_{2N_\rmf})_\rmA}{\mathbb{Z}_{N_\rmc}\times (\mathbb{Z}_{N_\rmf})_\rmL\times (\mathbb{Z}_{N_\rmf})_\rmR\times \mathbb{Z}_2}. 
\label{eq:symmetry_massless_QCD}
\ee
This is the correct global symmetry of massless QCD, in the sense that $G$ has the faithful representation on the physical Hilbert space\footnote{For any different $g,g'\in G$, there exists a gauge-invariant local operator $O(x)$ such that $g\cdot O(x)\not=g'\cdot O(x)$. }. 
For our purpose, it is an important step to identify the division by the discrete subgroup in (\ref{eq:symmetry_massless_QCD}) in a correct manner, and thus we will explain it in detail. 
There are several equivalent expressions of the symmetry group (\ref{eq:symmetry_massless_QCD}),  and each of them has pros and cons. Later in this section, we shall also discuss it. 

Since quark fields are massless, we can rotate left- and right-handed quarks, $\psi_{\rmL,\rmR}={1\mp \gamma_5\over 2}\psi$, separately. Quark fields form $N_\rmf$-dimensional complex vectors, and they are in the faithful representation of $U(N_\rmf)_\rmL\times U(N_\rmf)_\rmR$. When writing the unitary matrix as a product of a special unitary matrix and $U(1)$ phase factor, there is a redundancy related to the center element of the $SU(N_\rmf)$ matrix, so $U(N_\rmf)=[SU(N_\rmf)\times U(1)]/\mathbb{Z}_{N_\rmf}$. 
As a result, the flavor symmetry of the classical Lagrangian (\ref{eq:classical_action_QCD}) is 
\bea
G_{\mathrm{classical}}^{(\mathrm{quark})}&=&U(N_\rmf)_\rmL\times U(N_\rmf)_\rmR\nonumber\\
&=&{SU(N_\rmf)_\rmL\times U(1)_{\rmL}\over (\mathbb{Z}_{N_\rmf})_\rmL}\times {SU(N_\rmf)_\rmR\times U(1)_{\rmR}\over (\mathbb{Z}_{N_\rmf})_\rmR}. 
\eea
We can rewrite $U(1)_\rmL\times U(1)_\rmR$ by vector and axial $U(1)$ symmetries:
\be
\rme^{\im \alpha_\rmL(1-\gamma_5)/2} \rme^{\im \alpha_\rmR (1+\gamma_5)/ 2}= \rme^{\im (\alpha_\rmL+\alpha_\rmR)/2}\rme^{\im \gamma_5(\alpha_\rmR-\alpha_\rmL)/2}.
\ee
Here, we should notice that the angles of vector and axial $U(1)$ rotations are both divided by $2$. As a consequence, $U(1)_\rmL\times U(1)_\rmR=[U(1)_\rmV\times U(1)_\rmA]/\mathbb{Z}_2$, which can be understood by finding the $\pi$ rotations of the $U(1)_\rmV$ and $U(1)_\rmA$ symmetries are the same element; $\mathrm{e}^{\im \pi}=\mathrm{e}^{\im \pi\gamma_5}=-1$. The classical flavor symmetry is now written as 
\be
G_{\mathrm{classical}}^{(\mathrm{quark})}=\frac{SU(N_\rmf)_\rmL \times SU(N_\rmf)_\rmR\times U(1)_\rmV\times U(1)_\rmA}{(\mathbb{Z}_{N_\rmf})_\rmL\times (\mathbb{Z}_{N_\rmf})_\rmR\times \mathbb{Z}_2}. 
\ee
Now, we must take into account the effect of the fermion measure. Quantum mechanically, $U(1)_\rmA$ is explicitly broken, and the measure $\Diff \overline{\psi}\Diff \psi$ changes under the chiral transformation, $\psi\mapsto \exp(\im\alpha_\rmA \gamma_5)\psi$ and $\overline{\psi}\mapsto \overline{\psi}\exp(\im \alpha_\rmA \gamma_5)$, as 
\be
\Diff \overline{\psi}\Diff \psi\mapsto \Diff \overline{\psi} \Diff \psi \exp\left(2\im \alpha_\rmA{N_f\over 8\pi^2}\int \tr_\rmc(F_\rmc\wedge F_\rmc)\right). 
\ee
Since the topological charge $Q={1\over 8\pi^2}\int \tr_\rmc(F_\rmc\wedge F_\rmc)$ is always an integer on any orientable closed manifolds, the transformation is a symmetry only if $\alpha_\rmA$ is quantized to $2\pi/(2 N_\rmf)$. 
As a result, $G^{(\mathrm{quark})}$ is explicitly broken down to 
\be
G^{(\mathrm{quark})}=\frac{SU(N_\rmf)_\rmL \times SU(N_\rmf)_\rmR\times U(1)_\rmV\times (\mathbb{Z}_{2N_\rmf})_\rmA}{(\mathbb{Z}_{N_\rmf})_\rmL\times (\mathbb{Z}_{N_\rmf})_\rmR\times \mathbb{Z}_2}. 
\ee

$G^{(\mathrm{quark})}$ acts on the quark field $\psi$ faithfully, but the quark field is not gauge invariant. There is still redundancy in $G^{(\mathrm{quark})}$ on the physical Hilbert space. 
Since physical operators must be singlet under the $SU(N_\rmc)$ gauge group, the $U(1)_\rmV$ charges of any gauge-invariant local operators are quantized to $N_\rmc$ because of $N_c$-ality. For instance, gluon operator $\tr_\rmc(F_\rmc\wedge \star F_\rmc)$, meson field $\calM\sim \overline{\psi}\psi$, baryon field $\calB\sim \psi^{N_\rmc}$ have charge $0$, $0$, $N_\rmc$, respectively. Therefore, the vector rotation by $\rme^{2\pi \im/N_\rmc}$ must be regarded as the identity, and the faithful flavor symmetry has to be divided by $\mathbb{Z}_{N_\rmc}$. We obtain the physical symmetry group as
\be
G=G^{(\mathrm{quark})}/\mathbb{Z}_{N_\rmc}, 
\ee 
and this gives (\ref{eq:symmetry_massless_QCD}).

\subsection{Other equivalent expressions of the flavor symmetry}

There are several equivalent expressions of the symmetry group (\ref{eq:symmetry_massless_QCD}), and some of them makes its physical meaning more apparent. First, we explain that the expression (\ref{eq:symmetry_massless_QCD}) can be simplified as 
\be
G={SU(N_\rmf)_\rmL\times SU(N_\rmf)_\rmR\times U(1)_\rmV \over \mathbb{Z}_{N_\rmc}\times (\mathbb{Z}_{N_\rmf})_\rmV}. 
\ee
This is because the discrete axial symmetry $(\mathbb{Z}_{2N_\rmf})_{\rmA}$ is a subgroup of continuous axial symmetry, 
\be
(\mathbb{Z}_{2N_\rmf})_\rmA\subset SU(N_\rmf)_\rmL\times SU(N_\rmf)_\rmR\times U(1)_\rmV.
\ee 
Indeed, the discrete axial symmetry is generated by $\exp\left({2\pi\im\over 2N_\rmf}\gamma_5\right)$, but this generator can be written as 
\bea
\exp\left({2\pi\im\over 2N_\rmf}\gamma_5\right)&=&\exp\left({2\pi\im\over N_\rmf}{1+\gamma_5\over 2}\mathrm{diag}[1,\ldots,1,1-N_\rmf]\right)\cdot \exp\left(-{2\pi\im\over 2N_\rmf}\right)\nonumber\\
&\in& SU(N_\rmf)_\rmR\times U(1)_\rmV. 
\eea
This redundancy is already taken into account in (\ref{eq:symmetry_massless_QCD}), because the discrete axial rotation can always be canceled by the denominator of (\ref{eq:symmetry_massless_QCD}) so that $[U(1)_\rmV\times (\mathbb{Z}_{2N_\rmf})_{\rmA}]/[(\mathbb{Z}_{N_\rmf})_\rmL\times (\mathbb{Z}_{N_\rmf})_{\rmR}\times \mathbb{Z}_2]=U(1)_\rmV/\mathbb{Z}_{N_\rmf}$. 
This expression clarifies that the flavor symmetry group for the fundamental quark is connected. This is a special feature of the defining representation, and the symmetry group of a matter fields in a higher representation typically contains disconnected component related to the anomaly free subgroup of $U(1)$ axial symmetry. 
In this paper, we prefer to use (\ref{eq:symmetry_massless_QCD}) because the discrete axial symmetry plays an important role. 

We can further simplify the expression by introducing the $U(1)$ baryon symmetry by 
\be
U(1)_\rmB=U(1)_\rmV/\mathbb{Z}_{N_\rmc}. 
\ee
As we have explained, the physical local operator always have the charge in $N_\rmc \mathbb{Z}$ under $U(1)_\rmV$ because it defines the quark number. By changing the normalization of the generator, we can define the baryon charge, which is given by $U(1)_\rmB$. As a result, we obtain the flavor symmetry group as 
\be
G={SU(N_\rmf)_\rmL\times SU(N_\rmf)_\rmR\times U(1)_\rmB\over \mathbb{Z}_{N_\rmf}}. 
\label{eq:symmetry_QCD_physical}
\ee
This expression is the most useful expression when discussing the physical spectrum of massless QCD. On the other hands, quarks in the QCD Lagrangian have fractional charges with this representation. 
In order to discuss anomaly matching, we have to introduce the $G$-gauge field and examine its gauge invariance, but the existence of fractional charges makes this examination more difficult. 
For derivation of our new anomaly matching condition, the expression (\ref{eq:symmetry_massless_QCD}) turns out to be more useful. After derivation, we make connection with (\ref{eq:symmetry_QCD_physical}) to understand the physical meaning of our result. 

\subsection{Background gauge fields and two-form gauge fields}

In order to find the 't~Hooft anomaly of massless QCD, we introduce the background gauge field for the global symmetry $G$, and examine the gauge-invariance of the partition function. 
Because of the nontrivial topology of $G$, its gauge field consist not only of the one-form gauge field but also of the two-form gauge field. 
In this subsection, we explain why such unconventional gauge field appears by taking $U(1)_\rmV/\mathbb{Z}_{N_\rmc}\subset G$ as a simple example. Full description of $G$-gauge field will be given in Sec.~\ref{sec:anomaly}. 

First, let us describe the mathematical data of massless QCD before gauging $U(1)_\rmV/\mathbb{Z}_{N_\rmc}$. 
Massless QCD is an $SU(N_\rmc)$ gauge theory, so it is given by the principal $SU(N_\rmc)$ bundle with fundamental quarks. Therefore, we introduce the open cover $\{U_i\}$ of the Euclidean spacetime $M_4$. The $SU(N_\rmc)$ gauge field $a$ is the collection of $\mathfrak{su}(N_c)$-valued one-forms $a_i$ on $U_i$ and $SU(N_\rmc)$-valued transition functions $g^{\rmc}_{ij}$ on $U_{ij}=U_i\cap U_j$, with 
\be
a_j=(g^\rmc_{ij})^{-1}a_i g^\rmc_{ij}+(g^\rmc_{ij})^{-1}\diff g^\rmc_{ij}. 
\ee
We define $g^\rmc_{ji}=(g^\rmc_{ij})^{-1}$, and require the cocycle condition on the triple overlaps $U_{ijk}=U_{ij}\cap U_{jk}\cap U_{ki}$, 
\be
g^\rmc_{ij} g^\rmc_{jk} g^\rmc_{ki}=1. 
\label{eq:cocycle_SU(Nc)}
\ee
Whether or not the fundamental matter exists in the theory, we call the theory as an $SU(N_\rmc)$ gauge theory if the cocycle condition (\ref{eq:cocycle_SU(Nc)}) is satisfied. In our case, the quark field in the fundamental representation requires (\ref{eq:cocycle_SU(Nc)}) as a consistency condition: 
The quark field $\psi$ is also given as the collection of Grassmannian field $\psi_i$ on each open set $U_i$, with the connection formula 
\be
\psi_j=(g^\rmc_{ij})^{-1}\cdot \psi_i 
\ee
on the double overlaps $U_{ij}$. Uniqueness of $\psi_i$ on the triple overlap $U_{ijk}$ requires (\ref{eq:cocycle_SU(Nc)}). 

Let us perform the gauging of $U(1)_\rmV/\mathbb{Z}_{N_\rmc}$. Now, the gauge group becomes $[SU(N_\rmc)\times U(1)_\rmV]/\mathbb{Z}_{N_\rmc}$. The one-form gauge fields are $\mathfrak{su}(N_c)$-valued one-form $a_i$ and $\mathfrak{u}(1)$-valued one-form $A_{\rmV,i}$ on $U_i$ with the connection formula on double overlaps $U_{ij}$, 
\bea
a_j&=&(g^\rmc_{ij})^{-1}a_i g^\rmc_{ij}+(g^\rmc_{ij})^{-1}\diff g^\rmc_{ij}, \\
A_{\rmV,j}&=&(g^\rmV_{ij})^{-1}A_{\rmV,i} g^\rmV_{ij}+(g^\rmV_{ij})^{-1}\diff g^\rmV_{ij},
\eea
and $g^\rmc_{ij}$ and $g^\rmV_{ij}$ are $SU(N_\rmc)$- and $U(1)$-valued transition functions, respectively. 
Since the quark field $\psi$ is in the defining representation of $SU(N_\rmc)$ and has charge $1$ under $U(1)_\rmV$, its connection formula is given by 
\be
\psi_j =(g^\rmc_{ij}\cdot g^\rmV_{ij})^{-1} \psi_i. 
\ee
Now, the consistency on the triple overlap does not require the naive cocycle condition given in (\ref{eq:cocycle_SU(Nc)}). Instead, the consistency only requires 
\be
g^\rmc_{ij} g^\rmc_{jk} g^\rmc_{ki}=\exp\left({2\pi\im\over N_\rmc} n_{ijk}\right),\quad 
g^\rmV_{ij} g^\rmV_{jk} g^\rmV_{ki}=\exp\left(-{2\pi\im\over N_\rmc}n_{ijk}\right)
\label{eq:cocycle_U(Nc)}
\ee
with $n_{ijk}\in\mathbb{Z}_{N_\rmc}$. When setting $n_{ijk}\equiv 0$ mod $N_\rmc$, the gauge group becomes $SU(N_\rmc)\times U(1)_\rmV$, and it corresponds to gauging of $U(1)_\rmV$. However, such a requirement is too strong, because the violation of the cocycle condition for $g^\rmc_{ij}$ by the center $\mathbb{Z}_{N_\rmc}$ can be compensated by $g^\rmV_{ij}$ as in (\ref{eq:cocycle_U(Nc)}). 
We argue in the above that there is a freedom to introduce additional data $\{n_{ijk}\}$ since the global symmetry with the faithful representation on physical spectrum is $U(1)_\rmV/\mathbb{Z}_{N_\rmc}$. 

This additional data $\{n_{ijk}\}$ is specified by the $\mathbb{Z}_{N_\rmc}$ two-form gauge field $\in H^2(M_4,\mathbb{Z}_{N_\rmc})$. 
We can see this by noticing that consistency of (\ref{eq:cocycle_U(Nc)}) on the quadruple overlap $U_{ijk\ell}=U_{ijk}\cap U_{ij\ell}\cap U_{ik\ell}\cap U_{jk\ell}$ requires 
\be
n_{ijk}-n_{ij\ell}+n_{i k \ell}-n_{jk\ell}=0\bmod N_\rmc. 
\label{eq:closedness_twoform}
\ee
Redefinition of transition functions, $g^\rmc_{ij}\mapsto g^\rmc_{ij}\exp\left({2\pi\im\over N_\rmc}n_{ij}\right)$ and $g^\rmV_{ij}\mapsto g^\rmV_{ij}\exp\left(-{2\pi\im\over N_\rmc}n_{ij}\right)$, changes 
\be
n_{ijk}\mapsto n_{ijk}+n_{ij}+n_{jk}+n_{ki}, 
\ee
without affecting the connection formula of gauge fields and quark fields. Therefore, we can introduce the identification, $\{n_{ijk}\}\sim \{n_{ijk}+n_{ij}+n_{jk}+n_{ki}\}$. The equivalence class $[\{n_{ijk}\}]$ satisfying (\ref{eq:closedness_twoform}) is nothing but the mathematical definition of $B^{(2)}_\rmc\in H^2(M_4,\mathbb{Z}_{N_\rmc})$. 
The equivalence relation $\{n_{ijk}\}\sim \{n_{ijk}+n_{ij}+n_{jk}+n_{ki}\}$ says that the gauged theory is invariant under the $\mathbb{Z}_{N_\rmc}$ one-form gauge transformations.

There is a useful description for continuum field theories to take into account the effect of $B^{(2)}_\rmc\in H^2(M_4,\mathbb{Z}_{N_\rmc})$. Let us explain it by extending the discussion of Ref.~\cite{Kapustin:2014gua}. 
We can realize the $\mathbb{Z}_{N_\rmc}$ two-form gauge field as a pair of $U(1)$ two-form and one-form gauge fields, $(B^{(2)}_\rmc, B^{(1)}_\rmc)$, satisfying the constraint, 
\be
N_\rmc B^{(2)}_\rmc=\diff B^{(1)}_\rmc. 
\ee
We then consider the $U(N_\rmc)\times U(1)_\rmV$ gauge theory by embedding $SU(N_\rmc)$ connection $a$ into the $U(N_\rmc)$ connection: We describe the corresponding $U(N_\rmc)$ connection by $\widetilde{a}$, which is locally consist of $SU(N_\rmc)$ connection $a$ and $U(1)$ connection $B^{(1)}_\rmc$: 
\be
\widetilde{a}=a+{1\over N_\rmc}B^{(1)}_\rmc \bm{1}_{N_\rmc}. 
\label{eq:U(N_c)_local}
\ee
Since we are now considering the $U(N_\rmc)$ principal bundle instead of the $SU(N_\rmc)$ principal bundle, the gauge transformation is parametrized by $U(N_\rmc)$-valued functions $g_\rmc(x)\in U(N_\rmc)$ instead of $SU(N_\rmc)$-valued functions: 
\be
\widetilde{a}\mapsto g^\dagger_\rmc (\widetilde{a}+\diff) g_\rmc,\; \psi\mapsto g_\rmc^{\dagger} \psi, \; \overline{\psi}\mapsto \overline{\psi}g_\rmc. 
\ee
Therefore, each term on the right hand side of (\ref{eq:U(N_c)_local}), $a$ and ${1\over N_\rmc}B^{(1)}_\rmc$, does not have a gauge-invariant meaning globally. What (\ref{eq:U(N_c)_local}) implies is that the path integral $\int \Diff \widetilde{a}$ sums up all $U(N_\rmc)$ gauge connections, $\widetilde{a}$, satisfying 
\be
B^{(1)}_\rmc=\tr_\rmc[\widetilde{a}].
\ee 
Now, $\overline{\psi}\gamma_{\mu}D_{\mu}(\widetilde{a})\psi$ is invariant under this local $U(N_\rmc)$ gauge transformation. 
What we want to do is to put the theory on $[SU(N_\rmc)\times U(1)_\rmV]/\mathbb{Z}_{N_\rmc}$ principal bundles. For that purpose, we postulate the invariance under the $U(1)$ one-form gauge transformation~\cite{Kapustin:2014gua}, defined by 
\bea
&&B^{(2)}_\rmc\mapsto B^{(2)}_\rmc+\diff \lambda^{(1)}_\rmc,\; B^{(1)}_\rmc\mapsto B^{(1)}_\rmc+ N_\rmc \lambda^{(1)}_\rmc, \; \nonumber\\
&&\widetilde{a}\mapsto \widetilde{a}+\lambda^{(1)}_\rmc,\; 
A_\rmV\mapsto A_\rmV-\lambda^{(1)}_c,
\label{eq:Z_Nc_gauge_transformation}
\eea
where $\lambda_\rmc$ is the gauge parameter and the $U(1)$ gauge field. The transformation law for $\widetilde{a}$ is determined so that it is consistent with the local expression (\ref{eq:U(N_c)_local}). 
The transformation law for $U(1)_\rmV$ is chosen so that the covariant derivative with the $U(1)_\rmV$ gauge field, 
\be
D(\widetilde{a},A_\rmV)\psi=\left(\diff+\widetilde{a}+A_{\rmV}\right)\psi, 
\ee
is invariant under this one-form gauge transformation. 
By keeping the invariance under this transformation (\ref{eq:Z_Nc_gauge_transformation}), we can eliminate the double counting of the gauge group elements. 

Before going to the explanation on other background gauge fields, we still need to look at the invariance of the kinetic term $\tr_\rmc(F_\rmc\wedge \star F_\rmc)$ under the $U(N_\rmc)$ gauge transformation and $U(1)$ one-form gauge transformation. 
In order to make it invariant under the $U(N_\rmc)$ gauge transformation, we need to replace $F_\rmc(a)$ by 
\be
\widetilde{F}_\rmc\equiv F_\rmc(\widetilde{a})=\diff \widetilde{a}+\widetilde{a}\wedge \widetilde{a}. 
\ee
Under the $U(N_\rmc)$ gauge transformation (\ref{eq:U(N_c)_local}), this field strength changes as $F_\rmc(\widetilde{a})\mapsto gF_\rmc(\widetilde{a})g^{\dagger}$, and thus the kinetic term becomes invariant. 
Under the $U(1)$ one-form gauge transformation (\ref{eq:Z_Nc_gauge_transformation}), it changes as 
\be
F_\rmc(\widetilde{a})\mapsto F_\rmc(\widetilde{a})+\diff \lambda^{(1)}_\rmc. 
\ee
The gauge invariant combination is $F_\rmc(\widetilde{a})-B^{(2)}_\rmc$, and thus the gauge-invariant kinetic term for the gauge field is now given by
\be
{1\over 2g^2}\int \tr_{\rmc}\bigl[(F_\rmc(\widetilde{a})-B^{(2)}_\rmc)\wedge \star (F_\rmc(\widetilde{a})-B^{(2)}_\rmc)\bigr]. 
\ee
In order to judge whether we obtain the $[SU(N_\rmc)\times U(1)_\rmV]/\mathbb{Z}_{N_\rmc}$ gauge theory by this procedure, we should look at the spectrum of the gauge-invariant genuine line operators~\cite{Kapustin:2014gua}: All the genuine line operators must have zero charge under the $\mathbb{Z}_{N_\rmc}$ one-form transformation. 
In our case, the $U(1)$ one-form gauge invariance (\ref{eq:Z_Nc_gauge_transformation}) establishes this condition, and examples of the gauge-invariant line genuine operators along the closed line $L$ are $\mathrm{tr}_\rmc\left[\mathcal{P}\exp(\int_L \widetilde{a})\right]\exp(\int_L A_\rmV)$, $\left(\mathrm{tr}_\rmc\left[\mathcal{P}\exp(\int_L \widetilde{a})\right]\right)^{N_\rmc}\exp(-\int_L B^{(1)}_\rmc)$, and so on, which have indeed charge zero under the diagonal center, $\mathbb{Z}_{N_\rmc}\subset SU(N_\rmc)\times U(1)_\rmV$. 

The two-form gauge field $B^{(2)}_\rmc$ has a striking effect on the topological charge~\cite{vanBaal:1982ag}. The index theorem of the Dirac operator in the defining representation tells that the $SU(N)$ topological charge is quantized to integers, 
\be
{1\over 8\pi^2}\int_{M_4}\tr_\rmc[F_\rmc\wedge F_\rmc]\in \mathbb{Z}, 
\ee
but it becomes fractional after introducing $B^{(2)}_\rmc$: 
\be
{1\over 8\pi^2}\int_{M_4}\tr_\rmc \left[\left(\widetilde{F}_\rmc-B^{(2)}_\rmc\right)^2\right]\in {1\over N_\rmc}\mathbb{Z}. 
\ee 
To see it, we expand the left hand side while the expansion hides the manifest one-form gauge invariance, and we get 
\bea
{1\over 8\pi^2}\int_{M_4}\tr_\rmc \left[\left(\widetilde{F}_\rmc-B^{(2)}_\rmc\right)^2\right]
&=&{1\over 8\pi^2}\int_{M_4}\left(\tr_\rmc\left[\widetilde{F}_\rmc^2\right]-2 \tr_\rmc \left[\widetilde{F}_\rmc\right]\wedge B_\rmc^{(2)}+N_\rmc B^{(2)}_\rmc\wedge B^{(2)}_\rmc\right)\nonumber\\
&=&{1\over 8\pi^2}\int_{M_4}\tr_\rmc\left[\widetilde{F}_\rmc^2\right]-{N_\rmc\over 8\pi^2}\int_{M_4} B^{(2)}_\rmc\wedge B^{(2)}_\rmc. 
\eea
The first term on the right-hand-side is the topological charge of the $U(N_\rmc)$ gauge field strength, $\widetilde{F}_\rmc$, and gives an integer. The second term is quantized to $1/N_\rmc$, because $N_\rmc B^{(2)}_\rmc=\diff B^{(1)}_\rmc$, and we get the result. A more explicit proof on hypertorus is given in \cite{vanBaal:1982ag}. 

Lastly, it would be useful to discuss the baryon charge $U(1)_\rmB$. Since the baryon operator $\calB\sim \psi^{N_\rmc}$ has the charge $N_\rmc$ of $U(1)_\rmV$, the covariant derivative on it should look like $D\calB\sim (\diff+N_\rmc A_\rmV)\calB$. However, this derivative is not invariant under the $U(1)$ one-form gauge transformation, so the correct one must be 
\be
D\calB=(\diff + N_\rmc A_\rmV+B^{(1)}_\rmc)\calB.
\ee 
This tells us that the gauge field $A_\rmB$ of $U(1)_\rmB$ is identified as 
\be
A_\rmB=N_\rmc A_\rmV+B^{(1)}_\rmc. 
\label{eq:gauge_field_baryon}
\ee
Its field strength is given as 
\be
\diff A_\rmB=N_\rmc \diff A_\rmV+\diff B^{(1)}_\rmc=N_\rmc (\diff A^{(1)}_\rmV+B^{(2)}_\rmc). 
\ee
Because of the appearance of $\diff B^{(1)}_\rmc(=N_\rmc B^{(2)}_\rmc)$ in this expression, $A_\rmB$ is canonically normalized as a $U(1)$ gauge field:
\be
{1\over 2\pi}\int \diff A_\rmB={N_\rmc \over 2\pi}\int (\diff A_\rmV+B^{(2)}_\rmc)\in\mathbb{Z}. 
\ee
Therefore, the $\mathbb{Z}_{N_\rmc}$ two-form gauge field $(B^{(2)}_\rmc, B^{(1)}_\rmc)$ has an important physical meaning in massless QCD. 
\section{Discrete 't Hooft anomaly of massless QCD}\label{sec:anomaly}

't~Hooft anomaly of the global symmetry $G$ is defined by the absence of $G$-gauge invariance when the $G$-background gauge fields are introduced. 
To detect the anomaly, we need introduce the $G$-gauge field for (\ref{eq:symmetry_massless_QCD}). In order to emphasize the role of discrete axial symmetry $(\mathbb{Z}_{2N_\rmf})_\rmA$, we will consider a subgroup $G^{\mathrm{sub}}\subset G$ and introduce the $G^{\mathrm{sub}}$-gauge field instead. After that, we examine the gauge invariance of the partition function and derive the 't~Hooft anomaly by using the Stora-Zumino descent procedure.

\subsection{Background gauge fields and UV regularization of quark fields}

We introduce the $G$-gauge field in order to detect the 't~Hooft anomaly. In this paper, we would like to clarify the role of the discrete axial symmetry $(\mathbb{Z}_{2N_\rmf})_\rmA$, and for that purpose we especially consider the subgroup, 
\be
G^\mathrm{sub}\equiv{SU(N_\rmf)_\rmV\times U(1)_\rmV \times (\mathbb{Z}_{2N_\rmf})_\rmA\over \mathbb{Z}_{N_\rmc}\times (\mathbb{Z}_{N_\rmf})_\rmV\times \mathbb{Z}_2} \subset G. 
\ee
It is useful to rewrite $G^\mathrm{sub}$ as 
\be
G^{\mathrm{sub}}={SU(N_\rmf)_\rmV\times U(1)_\rmV \over \mathbb{Z}_{N_\rmc}\times (\mathbb{Z}_{N_\rmf})_\rmV}\times (\mathbb{Z}_{N_\rmf})_\rmL, 
\ee
since this expression has less redundancy and simplifies the computation. 
The background $G^{\mathrm{sub}}$ gauge field consists of 
\begin{itemize}
\item $A_\rmf$: $SU(N_\rmf)_\rmV$ one-form gauge field, 
\item $A_\rmV$: $U(1)_\rmV$ one-form gauge field, 
\item $A^{(1)}_\chi$: $(\mathbb{Z}_{N_\rmf})_\rmL$ one-form gauge field, 
\item $B^{(2)}_\rmc$: $\mathbb{Z}_{N_\rmc}$ two-form gauge field, 
\item $B^{(2)}_\rmf$: $(\mathbb{Z}_{N_\rmf})_\rmV$ two-form gauge field.  
\end{itemize}
As we have done in the previous section, we realize the $\mathbb{Z}_N$ $p$-form gauge field $A^{(p)}$ as a pair $(A^{(p)}, A^{(p-1)})$ of $U(1)$ $p$-form and $(p-1)$-form gauge fields, satisfying the constraint $N A^{(p)}=\diff A^{(p-1)}$~\cite{Banks:2010zn}
\footnote{Since $A^{(p-1)}$ is the phase function of $(p-1)$-form Higgs field, we can introduce some external flux by defects, at which the Higgs vacuum expectation value disappears, and $A^{(p)}$ should be regarded as an almost flat connection but not completely flat (see Ref.~\cite{Wang:2014pma}). In this paper, it is enough to know that $\diff A^{(1)}_\chi\not=0$ and it still satisfies the quantization $\int A^{(1)}_\chi\in {2\pi\over N_\rmf}\mathbb{Z}$ on closed manifolds. For more details on the mathematical side, see Ref.~\cite{Kapustin:2014zva}.}. 

We embed the $SU(N_{\rmc,\rmf})$ gauge fields into $U(N_{\rmc,\rmf})$ gauge fields, locally given as 
\be
\widetilde{a}=a+{1\over N_\rmc}B^{(1)}_{\rmc}\bm{1}_{N_\rmc},\; \widetilde{A}_{\rmf}=A_\rmf+{1\over N_\rmf}B^{(1)}_\rmf \bm{1}_{N_\rmf}. 
\ee
In order to describe the correct quotient of $G^{\mathrm{sub}}$, we must postulate the invariance under one-form gauge transformations,
\bea
&&B^{(2)}_\rmc \mapsto B^{(2)}_{\rmc}+ \diff \lambda^{(1)}_\rmc,\; B^{(1)}_\rmc\mapsto B^{(1)}_\rmc+N_\rmc \lambda^{(1)}_\rmc,\\
&&B^{(2)}_\rmf \mapsto  B^{(2)}_{\rmf}+\diff \lambda^{(1)}_\rmf,\; B^{(1)}_\rmf\mapsto B^{(1)}_\rmf+N_\rmf \lambda^{(1)}_\rmf, 
\eea
where the gauge parameter $\lambda^{(1)}_{\rmc,\rmf}$ are canonically normalized $U(1)$ gauge fields. 
The ordinary gauge fields transform under this one-form symmetry as 
\bea
\widetilde{a}&\mapsto& \widetilde{a}+ \lambda^{(1)}_\rmc \bm{1}_{N_\rmc},\\
\widetilde{A}_{\rmf} &\mapsto & \widetilde{A}_{\rmf}+ \lambda^{(1)}_\rmf \bm{1}_{N_\rmf},\\
A_\rmV&\mapsto& A_\rmV-\lambda^{(1)}_\rmc-\lambda^{(1)}_\rmf, 
\eea
and $A^{(1)}_\chi$ is invariant under one-form symmetry.

The quark kinetic term is now given by 
\bea
\overline{\psi}\gamma_\mu {D}_\mu \psi
&=&\overline{\psi}\gamma_\mu \left(\p_\mu+[\widetilde{a}+\widetilde{A}_\rmf+A_\rmV+A^{(1)}_\chi]_{\mu}\right) P_\rmL\psi\nonumber\\
&&+\overline{\psi}\gamma_{\mu} \left(\p_{\mu}+[\widetilde{a}+\widetilde{A}_\rmf+A_\rmV]_{\mu}\right) P_\rmR \psi, 
\eea
where $P_{\rmL,\rmR}=(1\mp \gamma_5)/2$ are chiral projectors. 
Since the Dirac operator is chiral, there is a possibility for the chiral anomaly. For computation of the anomaly, it is useful to rely on the fact that the above chiral Dirac operator is manifestly invariant under one-form gauge symmetry. 
Therefore, if we show that we can regularize the theory that keeps this manifest invariance under one-form gauge symmetry, then all we have to do is to use the standard technique for computing the non-Abelian consistent anomaly\footnote{When we are only interested in the anomaly linear in $A_\chi^{(1)}$, we can use the knowledge of Abelian anomaly after introducing the gauge fields of vector-like symmetries. Since this is a useful check of the result below, we give its result in the appendix~\ref{app:alternative_derivation}. }. Let us show that this is indeed the case. 

Following Sec.~3 of Ref.~\cite{AlvarezGaume:1984dr}\footnote{There is a possibility that a more subtle anomaly exists that cannot be captured by this procedure. To get it, one needs to introduce a five dimensional space $X$ so that $\p X=M_4$, and introduce the bundle structure on $X$ whose restriction to $M_4$ gives four-dimensional chiral Dirac operators (see Refs.~\cite{Witten:2015aba, Yonekura:2016wuc}). }, we first double the number of quark fields $\psi_\ell,\psi_r$, and replace the quark kinetic term as 
\bea
\overline{\psi}\gamma_{\mu}{D}_{\mu}\psi&\Rightarrow &
\overline{\psi_\ell}\gamma_{\mu}\left(\p_{\mu}+[\widetilde{a}+\widetilde{A}_\rmf+A_\rmV+A^{(1)}_\chi]_{\mu}P_{\rmL}\right)\psi_\ell\nonumber\\
&&+\overline{\psi_r}\gamma_{\mu}\left(\p_{\mu}+[\widetilde{a}+\widetilde{A}_\rmf+A_\rmV]_{\mu}P_{\rmR}\right)\psi_r, 
\eea
so that we have a well-defined eigenvalue problem for each left- and right-Dirac operators. 
Using the eigenvalues of this new left- and right-Dirac operators, we regularize the fermionic path-integral measure $\Diff\overline{\psi_\ell}\Diff\psi_\ell \Diff\overline{\psi_r}\Diff\psi_r$ by using the Fujikawa method~\cite{Fujikawa:1979ay, Fujikawa:1980eg, Fujikawa:2004cx}. 
At each step of the above ultraviolet regularization, we keep the manifest invariance under the one-form symmetry.  Under this regularization scheme, it is known that the anomaly of ordinary gauge symmetry is given by the consistent anomaly~\cite{AlvarezGaume:1984dr}, and we will compute below. 

\subsection{Computation of anomaly via Stora-Zumino chain}

The most convenient way to compute the consistent anomaly is to solve the Wess-Zumino consistency condition by using the descent procedure. 
We shall show the $5$-dimensional SPT action of the Stora-Zumino chain~\cite{Stora:1976kd, Stora:1983ct, Zumino:1983ew, AlvarezGaume:1984dr} is given by
\bea
S_{\mathrm{SPT}}&=&{N_\rmc\over 8\pi^2}\int  A^{(1)}_\chi\wedge  \tr_\rmf \left[\widetilde{F}_\rmf^2\right]+
{N_\rmf\over (2\pi)^2}\int A^{(1)}_\chi\wedge \diff A_\rmB \wedge B^{(2)}_f\nonumber\\
&\in&{2\pi\over N_\rmf}\mathbb{Z}.  
\label{eq:new_anomaly}
\eea
Here, we introduce the $U(1)_\rmB$ gauge field $A_\rmB$ by (\ref{eq:gauge_field_baryon}). 
Since this topological action is nontrivial mod $2\pi$, it gives 't~Hooft anomaly matching condition. 
The first term says that there is a mixed anomaly between $SU(N_\rmf)_\rmV$ and $(\mathbb{Z}_{2N_\rmf})_\rmA$ when $N_\rmf\not=N_\rmc$ , and the second term says that there is a mixed anomaly between $[SU(N_\rmf)\times U(1)_\rmB]/(\mathbb{Z}_{N_\rmf})_\rmV$ and $(\mathbb{Z}_{2N_\rmf})_\rmA$. 

We now derive (\ref{eq:new_anomaly}). The Stora-Zumino chain starts from the $6$-dimensional Abelian anomaly $\calA_6$, which is given by
\be
\calA_6={2\pi\over 3! (2\pi)^3}\int \tr_{\rmc,\rmf}\left[(\diff \calL+\calL^2)^3-(\diff \calR+\calR^2)^3\right], 
\ee
where $\calL$ and $\calR$ are the gauge fields coupled to left-handed and right-handed quarks, 
\be
\calL=\widetilde{a}+\widetilde{A}_\rmf+A_\rmV+A^{(1)}_\chi,\; \calR=\widetilde{a}+\widetilde{A}_\rmf+A_\rmV. 
\ee
Since $\calL=\calR+A^{(1)}_\chi$, we get 
\be
\calA_6={1\over 8\pi^2}\int \diff A^{(1)}_\chi \wedge \tr_{\rmc,\rmf}\left[(\diff\calR+\calR^2)^2\right] 
+O\left((\diff A_\chi^{(1)})^2\right). 
\ee
In this paper, we only pay attention to the anomaly polynomial linear in $A^{(1)}_\chi$ and neglect higher order terms, but it is straightforward to compute them. We shall see that the linear term in $A^{(1)}_\chi$ already gives interesting consequences on possible low-energy dynamics of massless QCD.  The descent procedure says that the $5$-dimensional parity anomaly is given by the boundary term of $\calA_6$, and it defines the $5$-dimensional topological action $S_{\mathrm{SPT}}$:
\bea
S_{\mathrm{SPT}}&=&{1\over 8\pi^2}\int A^{(1)}_\chi \tr_{\rmc,\rmf}[(\diff \calR+\calR^2)^2]\nonumber\\
&=&{1\over 8\pi^2}\int N_\rmf A^{(1)}_\chi \wedge \left(\tr_\rmc\left[\widetilde{F}_\rmc^2\right]+N_\rmc (\diff A_\rmV)^2\right)
+{N_\rmc \over 8\pi^2}\int A^{(1)}_\chi \wedge \tr_\rmf\left[\widetilde{F}_{\rmf}^2\right]\nonumber\\
&&+{1\over (2\pi)^2}\int A^{(1)}_\chi \wedge \left(N_\rmc \diff A_\rmV+\tr_\rmc\left[\widetilde{F}_\rmc\right]\right)\wedge \tr_{\rmf}\left[\widetilde{F}_\rmf\right]\nonumber\\
&&+{1\over (2\pi)^2}\int N_\rmf A^{(1)}_\chi \wedge  \tr_\rmc\left[\widetilde{F}_\rmc\right] \wedge \diff A_\rmV.
\label{eq:new_anomaly_1}
\eea
By construction, $S_{\mathrm{SPT}}$ is manifestly invariant under the one-form gauge transformation, and the Stora-Zumino procedure says that the gauge-dependence of the boundary term cancels the 't~Hooft gauge anomaly of massless QCD. 
Recalling that $N_\rmf A^{(1)}_\chi=\diff A^{(0)}_\chi$, the first and the last term of (\ref{eq:new_anomaly_1}) vanish modulo $2\pi$, and thus we can drop them. 
Let us demonstrate it for some of them:
\bea
{1\over 8\pi^2}\int N_\rmf A^{(1)}_\chi\wedge \tr_\rmc\left[\widetilde{F}_\rmc^2\right]&=&2\pi \int {\diff A^{(0)}_\chi\over 2\pi}\wedge {1\over 8\pi^2}\tr_\rmc\left[\widetilde{F}_\rmc^2\right]\nonumber\\
&=& 0\; \bmod 2\pi,\\
{1\over (2\pi)^2}\int N_\rmf A^{(1)}_\chi\wedge \tr_\rmc[\widetilde{F}_\rmc]\wedge \diff A_\rmV&=&
2\pi\int {\diff A^{(0)}_\chi\over 2\pi}\wedge {\diff B^{(1)}_\rmc \over 2\pi}\wedge {\diff A_\rmV\over 2\pi}\nonumber\\
&=& 0 \; \bmod 2\pi. 
\eea
For the first one, we use the index theorem for $U(N_\rmc)$ gauge field strength, $\widetilde{F}_\rmc$. 
We  now obtain (\ref{eq:new_anomaly}) from (\ref{eq:new_anomaly_1}) by using the $U(1)_\rmB$ gauge field $A_\rmB$ introduced as (\ref{eq:gauge_field_baryon}). 

The second term of (\ref{eq:new_anomaly}) is very interesting because this anomaly contains the gauge field for baryon charge $A_\rmB$.  Therefore, in any zero-temperature phase of massless QCD, the baryon charge must be defined with only massless fields and it must act nontrivially on the Hilbert space of the low-energy effective theory.

\section{Anomaly matching in chiral-symmetry broken phases}\label{sec:anomaly_matching}

In this section, we discuss the consequence of anomaly matching when chiral symmetry breaking occurs. 
The ordinary perturbative chiral anomaly is matched by the Wess-Zumino term of the pion Lagrangian, but we shall see that (\ref{eq:new_anomaly}) contains a term that cannot be produced by the Wess-Zumino term. 
We will find that the nontrivial topology of the vacuum manifold plays an important role in order to match the anomaly, and this is exactly the reason why nucleons can be described as skyrmions. 

We also critically examine an exotic chiral-symmetry broken phase without quark bilinear condensate, and the naive Stern phase is ruled out by anomaly matching argument. 

\subsection{Chiral symmetry breaking and Skyrmions}

We consider the 't~Hooft anomaly matching condition in the chiral symmetry broken phase with quark bilinear condensate. The symmetry breaking pattern is 
\be
G={SU(N_\rmf)_\rmL\times SU(N_\rmf)_\rmR\times U(1)_\rmB\over \mathbb{Z}_{N_\rmf}}\to H={SU(N_\rmf)_\rmV\times U(1)_\rmB\over \mathbb{Z}_{N_\rmf}}. 
\ee
The low-energy effective theory is given by the non-linear sigma model of Nambu-Goldstone bosons, and the target space is given by the coset,
\be
G/H={SU(N_\rmf)_\rmL\times SU(N_\rmf)_\rmR\over SU(N_\rmf)_\rmV}\simeq SU(N_\rmf). 
\ee
The sigma model field is described as $U:M_4\to SU(N_\rmf)$, and $U=\exp(\im \pi^a T^a/f_\pi)$. The left- and right-rotations $(g_\rmL,g_\rmR)\in [SU(N_\rmf)_\rmL\times SU(N_\rmf)_\rmR]/(\mathbb{Z}_{N_\rmf})_\rmV$ act on $U$ as 
\be
(g_\rmL,g_\rmR):U\mapsto g_\rmL U g_\rmR^\dagger,   
\ee
so we have a rough correspondence $U_{ff'}\sim \sum_c(\overline{\psi_\rmL})_{c f}(\psi_\rmR)_{c f'}$ and the order parameter $\langle \overline{\psi}\psi\rangle$ is proportional to $\left\langle \tr[U]+\tr[U^\dagger]\right\rangle$. 
It is important to notice that $U(1)_\rmB$ acts trivially on $U$. 

The Lagrangian of this theory is 
\be
S={f_\pi^2\over 2}\int \tr_\rmf[(U^{-1}\diff U)\wedge \star (U^{-1}\diff U)] + N_\rmc \Gamma_{\mathrm{WZ}}[U], 
\ee
where the second term is the Wess-Zumino term and it is necessary to match the perturbative anomaly~\cite{Wess:1971yu, Witten:1983tw}:
\be
\Gamma_{\mathrm{WZ}}={1\over 240\pi^2}\int_{M_5} \tr_\rmf[(U^{-1}\diff U)^5], 
\ee
with $\p M_5=M_4$. The coefficient is defined so that $\Gamma_{\mathrm{WZ}}$ does not depend on the extension of $U$ to $M_5$ up to $2\pi \mathbb{Z}$. It is straightforward (but a bit lengthy) to check that the Wess-Zumino term matches not only the perturbative non-Abelian anomaly but also the first term of (\ref{eq:new_anomaly}). However, $N_\rmc\Gamma_{\mathrm{WZ}}$ cannot match the second term of (\ref{eq:new_anomaly}) because $U(1)_\rmB$ acts trivially on $U$, so this must not be the whole story. 

Our anomaly matching condition says that we must be able to construct baryons using the low-energy effective field theory of the pion field $U$. This is indeed possible thanks to the topologically stable configuration\footnote{The topological stability does not necessarily leads the energetic stability. Indeed, Derick's theorem shows that the topologically stable configuration is unstable against the scale transformation within the lowest chiral Lagrangian. One necessarily adds a higher order term to evade this energetic instability, and such a term is known in this context as the Skyrme term. }, and such topological solitons are called skyrmions~\cite{Skyrme:1961vq, Skyrme:1962vh}. 
Let us remind the basic facts about skyrmions: Since the target space of the nonlinear sigma model is $[SU(N)_\rmL\times SU(N)_\rmR]/SU(N)_\rmV$, the topologically stable configurations are characterized by 
\be
\pi_3\left(G/H\right)=\pi_3(SU(N))\simeq \mathbb{Z}. 
\ee
Correspondingly, there exists the $U(1)$ symmetry that characterizes this topological number, and its Noether current is given by 
\be
J_\rmB={1 \over 24\pi^2}\tr_\rmf[(U^{-1}\diff U)^3]. 
\ee
The conservation law can be  checked that $\diff J_\rmB=-{1\over 24\pi^2}\tr_\rmf[(U^{-1}\diff U)^4]=0$ because of anti-commutation of the wedge product. The coefficient is normalized so that $\int_{S^3} J_\rmB$ is quantized to integers. 
We identify this $U(1)$ symmetry as $U(1)_\rmB$, and when gauging this symmetry we add the minimal-coupling term, 
\be
\int_{M_4} A_\rmB\wedge J_\rmB, 
\label{eq:baryon_minimal_coupling}
\ee
to the Lagrangian. Indeed, it is pointed out by Witten~\cite{Witten:1983tw} that this is the necessary term in order to reproduce the $U(1)_\rmV$-$SU(2)_\rmL$-$SU(2)_\rmL$ triangle anomaly. 

Let us now gauge $G^{\mathrm{sub}}=[SU(N_\rmf)_\rmV\times U(1)_\rmB]/(\mathbb{Z}_{N_\rmf})_\rmV\times (\mathbb{Z}_{N_\rmf})_\rmL$, and explicitly check that (\ref{eq:baryon_minimal_coupling}) leads to the anomaly matching for the second term of (\ref{eq:new_anomaly}). 
Naively, we would like to define $J_B$ with the background field by ${1\over 24\pi^2}\tr_\rmf[(U^{-1}DU)^3]$ with the covariant derivative, but such current does not obey the conservation law. 
To minimize this violation so that it is independent of the pion fields $U$, we add the counter term and define the gauge-invariant current as (see, e.g., Ref.~\cite{Harvey:2007ca})
\be
J_\rmB[\widetilde{A}_\rmf, A^{(1)}_\chi]={1\over 24\pi^2}\tr_\rmf [(U^{-1}D U)^3]+{1\over 8\pi^2}\tr_\rmf[(U D U^{-1})(\widetilde{F}_\rmf+\diff A^{(1)}_\chi)-(U^{-1}DU) \widetilde{F}_\rmf], 
\ee
where the covariant derivative is given by
\be
U^{-1}DU=U^{-1}\left[\diff U+(\widetilde{A}_\rmf+A^{(1)}_\chi) U-U \widetilde{A}_\rmf\right]. 
\ee
Indeed, we find that 
\be
\diff J_\rmB[\widetilde{A}_\rmf, A^{(1)}_\chi]={N_\rmf\over (2\pi)^2}\diff A^{(1)}_\chi\wedge  B^{(2)}_\rmf. 
\label{eq:4D_consistency}
\ee
The easiest way to obtain this result would be to use the fact that $J_\rmB$ is the three-dimensional Wess-Zumino term. Since it solves the consistency condition, its derivative is given by the four-dimensional Abelian anomaly, $\diff J_\rmB={1\over 2! (2\pi)^2}\tr_\rmf\left[(\widetilde{F}_\rmf+\diff A_\chi)^2-\widetilde{F}_\rmf^2\right]$, which gives (\ref{eq:4D_consistency}).  
The gauge variation of (\ref{eq:baryon_minimal_coupling}) gives
\be
\delta \int_{M_4}A_\rmB\wedge J_\rmB=\int_{M_4}\diff \lambda_\rmB\wedge J_\rmB=-{N_\rmf \over (2\pi)^2}\int_{M_4} \lambda_\rmB\wedge \diff A^{(1)}_\chi\wedge  B^{(2)}_\rmf, 
\ee
and this is the 't~Hooft anomaly characterized by the $5$-dimensional topological action
\be
-{N_\rmf\over (2\pi)^2}\int_{M_5} A_\rmB\wedge \diff A^{(1)}_\chi\wedge B^{(2)}_\rmf. 
\ee
This topological action is nothing but the second term of (\ref{eq:new_anomaly}) after integration by parts. This shows that the new discrete anomaly of massless QCD is matched thanks to the nontrivial property of skyrmion charges under background gauge fields.

\subsection{Ruling out chiral symmetry breaking without quark bilinear condensate}

In this section, we examine an exotic scenario of the chiral-symmetry broken phase of QCD, proposed by Stern~\cite{Stern:1997ri,Stern:1998dy} based on symmetries and anomalies. 
In a previous study~\cite{Kogan:1998zc}, this phase has been ruled out based on QCD inequalities, so it cannot be realized as the QCD vacuum at the zero baryon density. However, QCD  inequality cannot be applied when there exists a sign problem in the path integral, so it is still an open problem whether it appears, for example, at non-zero baryon densities or non-zero theta angles. 
We will negatively answer this question for the naive Stern phase, and our argument applies to much wider class of theories since it relies only on symmetries and anomalies. 

The conventional order parameter of chiral symmetry breaking is the chiral condensate $\langle \overline{\psi}\psi\rangle=\langle \overline{\psi_\rmR}\psi_\rmL\rangle+\langle \overline{\psi_\rmL}\psi_\rmR\rangle$, and the pion decay constant $f_{\pi}$ is another important parameter. 
Stern pointed out that the condition $f_{\pi}\not=0$ does not necessarily require that $\langle \overline{\psi}\psi\rangle \not=0$, and suggested the exotic chiral-symmetry broken phase with $f_{\pi}\not=0$ and $\langle \overline{\psi}\psi\rangle=0$~\cite{Stern:1997ri}. 
The local order parameter for this phase is the four-quark condensate~\cite{Kogan:1998zc}, such as 
\be
\sum_{a=1}^{N_\rmf^2-1}\Bigl\langle (\overline{\psi}T^a_\rmf P_\rmL \psi)(\overline{\psi}T^a_\rmf P_\rmR\psi)\Bigr\rangle=\sum_{a=1}^{N_\rmf^2-1}\Bigl\langle (\overline{\psi_\rmR}T^a_\rmf \psi_\rmL)(\overline{\psi_\rmL}T^a_\rmf \psi_\rmR)\Bigr\rangle, 
\ee
where $T^a_\rmf$ is the generator of the flavor symmetry. Under $SU(N_\rmf)_\rmL\times SU(N_\rmf)_\rmR$, the condensate transforms as the bi-adjoint representation, $(\bm{N_\rmf^2-1},\bm{N_\rmf^2-1})$. This condensate, therefore, breaks the continuous axial symmetry but keeps the discrete axial symmetry, 
\be
G\to G^{\mathrm{sub}}. 
\ee
This gives a natural explanation why the quark-bilinear condensate vanishes under this chiral-symmetry broken phase, since $\overline{\psi}\psi\mapsto \mathrm{e}^{2\pi\im/N_\rmf}\overline{\psi}\psi$ under the discrete chiral transformation $(\mathbb{Z}_{2N_\rmf})_\rmA$. 

In the method of phenomenological Lagrangian, the low-energy effective theory is described by the nonlinear sigma model with the target space 
\be
G/G^{\mathrm{sub}}={SU(N_\rmf)_\rmL \times SU(N_\rmf)_\rmR\over SU(N_\rmf)_\rmV\times (\mathbb{Z}_{N_\rmf})_\rmL}\simeq {SU(N_\rmf)\over \mathbb{Z}_{N_\rmf} }. 
\ee
We can realize this $SU(N_\rmf)/\mathbb{Z}_{N_\rmf}$ nonlinear sigma model as the usual $SU(N_\rmf)$ nonlinear sigma model with $\mathbb{Z}_{N_\rmf}$ gauge symmetry, and thus the only thing we have to do is to promote the background gauge field $A^{(1)}_\chi$ to a dynamical $(\mathbb{Z}_{N_\rmf})_\rmL$ gauge field $a^{(1)}_\chi$. 
As a result, $U\sim \overline{\psi_\rmL}\psi_\rmR$ is no longer a gauge invariant operator. 
Then, gauge invariance says that $\tr[U]$ cannot have non-zero expectation values, and one must construct an operator, such as $\tr[T^a_\rmf U^\dagger]\tr[T^a_\rmf U]$. 

However, there are two problems about this Stern phase. The first is the mismatch of symmetry. In general, when dynamically gauging the $\mathbb{Z}_N$ $p$-form symmetry in $d$-dimensional field theories, gauged theories acquire the dual $\mathbb{Z}_N$ $(d-p-2)$-form symmetry. 
Therefore, the four-dimensional $SU(N_\rmf)/\mathbb{Z}_{N_\rmf}$ nonlinear sigma model does not have the $(\mathbb{Z}_{N_\rmf})_\rmL$ $0$-form symmetry but instead has the $\mathbb{Z}_{N_\rmf}$ $2$-form global symmetry, which does not exist in QCD. The appearance of this two-form symmetry is related to the fact that $\pi_1(SU(N_\rmf)/\mathbb{Z}_{N_\rmf})\simeq \mathbb{Z}_{N_\rmf}$, and thus there exist topologically stable $(2+1)$D walls with the $\mathbb{Z}_{N_\rmf}$ charge. 
This mismatch of symmetry itself does not lead inconsistency, because it is possible that all particles with nontrivial $(\mathbb{Z}_{N_\rmf})_\rmL$ charge have mass gap and new symmetry often emerges at low energies.  

The second one is mismatch of anomaly, which is a more serious problem. Since there does not exist $\mathbb{Z}_{N_\rmf}$ $0$-form symmetry, the anomaly given by $S_{\mathrm{SPT}}$ in (\ref{eq:new_anomaly}) does not exist. 
The theory instead has a mixed 't~Hooft anomaly between $SU(N_\rmf)_\rmV/\mathbb{Z}_{N_\rmf}$ and $U(1)_\rmB$, but it is not of Dijkgraaf-Witten type in the terminology of Refs.~\cite{Kapustin:2014zva}. That is, anomalous violation of $J_\rmB$ does not take the form (\ref{eq:4D_consistency}) but takes 
\be
\diff J_\rmB={N_\rmf\over (2\pi)^2}\diff a^{(1)}_\chi\wedge B^{(2)}_\rmf, 
\ee
where $a^{(1)}_\chi$ is the auxiliary dynamical gauge field introduced above, and this difference causes the mismatch of anomaly. 
Therefore, the new anomaly matching condition by (\ref{eq:new_anomaly}) rules out the Stern phase even at finite densities although it can match the anomaly matching of perturbative non-Abelian anomalies. 
It is remarkable that we are now able to constrain not only the infinitesimal part of the target manifold of the nonlinear realization but also its topology by anomaly matching.

We point out that our result is consistent with that of the previous study with QCD inequalities~\cite{Kogan:1998zc}. 
Furthermore, our result gives the nontrivial extension because the QCD inequality is valid only if the path integral measure is positive definite, but the anomaly matching does not care about the sign problem. 
Therefore, we now find that the naive Stern phase cannot appear for the finite-density zero-temperature QCD by considering the background manifold as $M_4=S^1\times M_3$ in order to introduce the chemical potential and taking the infinite volume limit for the zero-mode projection. 

A possible detour evading this mismatch is to add additional massless excitations like color-singlet chiral fermions charged under $(\mathbb{Z}_{N_\rmf})_\rmL$, topological order, etc. It is an interesting study to examine this possibility from our anomaly matching condition, but let us stop here in this paper. 
We just point out that one should carefully design the contents of additional massless fields so that it does not produce additional perturbative anomaly because it is already matched by the Wess-Zumino term of pion fields. 

\subsection{Discussions on the large-$N_\rmc$ limit}\label{sec:discussion}

In this section, we give discussions to combine our no-go theorem on chiral symmetry breaking without quark bilinear condensates and the large-$N_\rmc$ limit. 

In the large-$N_\rmc$ limit, Coleman and Witten~\cite{Coleman:1980mx} have shown that chiral symmetry breaking is given by the orthodox one, $G\to H$, under the following assumptions\footnote{In the large-$N_\rmc$ limit, anomalous breaking $U(1)_\rmA\to (\mathbb{Z}_{2N_\rmf})_\rmA$ is subleading in the $1/N_\rmc$ expansion, so it is more correct to write $U(N_\rmf)_\rmL\times U(N_\rmf)_\rmR\to U(N_\rmf)_\rmV$. }:
\begin{enumerate}
\item Existence of asymptotic $1/N_\rmc$ expansions.
\item QCD shows confinement in the large-$N_\rmc$ limit.
\item Order parameter of chiral symmetry breaking is a quark-bilinear operator.
\item The ground state is the minimum of the effective potential $V$ of the order parameter.
\item No accidental symmetry appears in the effective potential $V$.  
\end{enumerate}

Let us now consider whether the discrete axial symmetry can be unbroken in the large-$N_\rmc$ QCD. 
To study such a possibility, we must replace the assumption $3$:
Instead of assuming that the order parameter is the quark-bilinear operator $U$ that transforms as $(g_\rmL,g_\rmR): U\to g_\rmL U g_\rmR^\dagger$, let us consider the case that the order parameter is given by the quark-quartic operator, $O$, that transforms as 
\be
O\mapsto (\mathrm{Ad}(g_\rmL)) O (\mathrm{Ad}(g_\rmR))^{-1}. 
\ee
We repeat the logic of Ref.~\cite{Coleman:1980mx} with this assumption. 
The effective potential is given by the invariant functional of $O$. Furthermore, the large-$N_\rmc$ counting shows that it must be an operator with the minimal number of trace\footnote{The operator $O$ is in the adjoint representation, so the minimal trace operation is given by the adjoint trace, and it corresponds to the double trace in the defining representation. Since we have assumed that the quark bilinear operator vanishes, the $1/N_\rmc$ expansion starts from the subleading order.}, so the possibility is 
\be
V(O)= \mathrm{Tr}(F(O^\dagger O)), 
\ee
where $F$ is an $N_\rmc$-independent function. This says that the minimum $O_*$ can be conjugated to a matrix $\lambda_*\bm{1}$. The anomaly matching for perturbative chiral anomaly shows that $\lambda_*\not=0$~\cite{Coleman:1980mx}, and then the chiral-symmetry breaking pattern is given by $G\to G^{\mathrm{sub}}$. 
We have however shown that this symmetry breaking is inconsistent with anomaly, at least at finite $N_\rmc$. Since we are assuming the smoothness of the large-$N_\rmc$ limit, the unbroken $(\mathbb{Z}_{N_\rmf})_\rmL$ symmetry is ruled out from the large-$N_\rmc$ QCD. 

This gives the argument why we can take the quark bilinear operator as an order parameter for the chiral symmetry breaking of large-$N_\rmc$ QCD, so a part of the assumptions in \cite{Coleman:1980mx} can be shown by discrete anomaly matching. 
We therefore conclude that our no-go theorem is nicely consistent with the theorem of chiral symmetry breaking in the large-$N_\rmc$ limit. 
\section{Consistency check for Seiberg duality of SUSY QCD}\label{sec:seiberg_duality}

In this section, we wish to discuss consistency of our anomaly matching condition with Seiberg duality of $\mathcal{N}=1$ supersymmetric QCD  (SQCD) with $N_\rmf\ge N_\rmc+2$~\cite{Seiberg:1994pq, Intriligator:1995au}. We also discuss the anomaly matching for $s$-confining phase at $N_\rmf=N_\rmc+1$. 

The $\mathcal{N}=1$ SQCD is the $\mathcal{N}=1$ $SU(N_\rmc)$ super Yang-Mills theory coupled to $N_\rmf$ flavor of chiral multiplet $Q$ in the $\bm{N_\rmc}$ representation and $\widetilde{Q}$ in the $\overline{\bm{N_\rmc}}$ representation. In addition to the global symmetry $G$ discussed in this paper, $\mathcal{N}=1$ SQCD also has the $U(1)$ $R$-symmetry, but we will not discuss it here; the charges are assigned so that only the $U(1)_\rmA$ symmetry is broken by quantum anomaly and becomes $(\mathbb{Z}_{2N_\rmf})_\rmA$.  The list of charge is summarized as 
\bea
\begin{array}{c| c | c | c| c | c}
 	&   SU(N_\rmc)   & SU(N_\rmf)_\rmL &  SU(N_\rmf)_\rmR  &  U(1)_\rmB  & (\mathbb{Z}_{N_\rmf})_\rmL\\ 
\hline
Q   &  \bm{N_\rmc}    & \bm{N_\rmf}        &    \bm{1}                &  1/N_\rmc & 1 \\
\widetilde{Q} & \overline{\bm{N_\rmc}} & \bm{1} &\overline{\bm{N_\rmf}} & -1/N_\rmc & 0
\end{array}
\eea
For $N_\rmf\ge N_\rmc+2$, $\mathcal{N}=1$ Seiberg duality proposes that this theory is dual to the $SU(N_\rmf-N_\rmc)$ gauge theory with the matter content,
\bea
\begin{array}{c| c | c | c| c | c}
 	&   SU(N_\rmf-N_\rmc)   & SU(N_\rmf)_\rmL &  SU(N_\rmf)_\rmR  &  U(1)_\rmB & (\mathbb{Z}_{N_\rmf})_\rmL \\ 
\hline
q   &  \bm{N_\rmf-N_\rmc}    & \overline{\bm{N_\rmf}}        &    \bm{1}                &  1/(N_\rmf-N_\rmc)  & -1\\ 
\widetilde{q} & \overline{\bm{N_\rmf-\bm{N}_\rmc}} & \bm{1} &\bm{N_\rmf} & -1/(N_\rmf-N_\rmc) & 0
\end{array}
\eea
Therefore, under this duality, $N_\rmc$ is mapped to $N_\rmf-N_\rmc$ and the chiral and flavor gauge fields are charge conjugated.

Let us check how $S_{\mathrm{SPT}}$ in (\ref{eq:new_anomaly}) is affected under this duality map. It is changed as 
\bea
S_{\mathrm{SPT}}&=&
{N_\rmc\over 8\pi^2}\int  A^{(1)}_\chi\wedge  \tr_\rmf \left[\widetilde{F}_\rmf^2\right]+
{N_\rmf\over (2\pi)^2}\int A^{(1)}_\chi\wedge \diff A_\rmB \wedge B^{(2)}_f\nonumber\\
&\mapsto&
{N_\rmf-N_\rmc\over 8\pi^2}\int  (-A^{(1)}_\chi)\wedge  \tr_\rmf \left[(-\widetilde{F}_\rmf^t)^2\right]+
{N_\rmf \over (2\pi)^2}\int (-A^{(1)}_\chi)\wedge \diff A_\rmB \wedge (-B^{(2)}_f)\nonumber\\
&=&S_{\mathrm{SPT}}-{N_\rmf\over 8\pi^2}\int A^{(1)}_\chi\wedge \tr_\rmf \left[\widetilde{F}_\rmf^2\right]\nonumber\\
&=&S_{\mathrm{SPT}} \; \bmod 2\pi. 
\eea
Therefore, we have confirmed that the anomaly matching of (\ref{eq:new_anomaly}) is also satisfied in  Seiberg duality of $\mathcal{N}=1$ $SU(N_\rmc)$ SQCD. 
This adds the additional evidence for the validity of Seiberg duality. This has been checked in a previous study~\cite{Shimizu:2017asf} but the discrete anomaly there exists only if $\mathrm{gcd}(N_\rmc,N_\rmf)>1$.  
We have extended the discrete anomaly for generic $N_\rmc$ and $N_\rmf$, so Seiberg duality passes more severe test. 

It is also interesting to consider the case $N_\rmf=N_\rmc+1$, at which the $s$-confining occurs. 
The $s$-confinement is the confinement without chiral symmetry breaking, and the dual theory is described only by gauge-singlet particles, mesons $\mathcal{M}$ and chiral baryons $\mathcal{B},\widetilde{\mathcal{B}}$:
\bea
\begin{array}{c| c | c| c | c}
 	&   SU(N_\rmf)_\rmL &  SU(N_\rmf)_\rmR  &  U(1)_\rmB  & (\mathbb{Z}_{N_\rmf})_\rmL\\ 
\hline
\mathcal{M}  &  \bm{N_\rmf}          &     \overline{\bm{N_\rmf}} &   0       &   1\\
\mathcal{B}        & \overline{\bm{N_\rmf}}        &    \bm{1}                &  1 & -1 \\
\widetilde{\mathcal{B}} & \bm{1} &{\bm{N_\rmf}} & -1 & 0
\end{array}
\eea
Let us check that the massless chiral baryons $\calB, \widetilde{\calB}$ produces the correct mixed anomaly involving $U(1)_\rmB$ and $(\mathbb{Z}_{N_\rmf})_\rmL$.  The covariant derivative acting on $\calB$ with the background gauge field is 
\bea
D \calB&=&(\diff - \widetilde{A}_\rmf + A_\rmB+B^{(1)}_\rmf-A^{(1)}_\chi) \calB, \\
D \widetilde{\calB}&=&(\diff + \widetilde{A}_\rmf - A_\rmB - B^{(1)}_\rmf) \widetilde{\calB}. 
\eea
We introduce the auxiliary gauge field $B^{(1)}_\rmf$ in order to make these covariant derivatives invariant under one-form gauge transformations. 
This produces the anomaly polynomial 
\bea
&&{1\over (2\pi)^2}\int (-A^{(1)}_\chi)\wedge \diff A_\rmB\wedge \tr_\rmf\left[-\widetilde{F}_\rmf+\diff B^{(1)}_\rmf\right]\nonumber\\
&=&{N_\rmf-N_\rmf^2 \over (2\pi)^2}\int A^{(1)}_\chi\wedge \diff A_\rmB\wedge B^{(2)}_\rmf\nonumber\\
&=&{N_\rmf \over (2\pi)^2}\int A^{(1)}_\chi\wedge \diff A_\rmB\wedge B^{(2)}_\rmf \; \bmod 2\pi. 
\eea
Therefore, the chiral baryons correctly reproduce the new anomaly matching condition in the $s$-confining phase. 
\section{Conclusion}\label{sec:conclusion}

In this paper, we derive the 't~Hooft anomaly of massless QCD that involves $U(1)_\rmB$ gauge field, $(\mathbb{Z}_{N_\rmf})_\rmL$ gauge field, and $(\mathbb{Z}_{N_\rmf})$ two-form gauge field. 
In order to properly gauge the symmetry, we have to introduce both one-form and two-form gauge fields in order to take into account the quotient structure of the symmetry group. 
This is important to obtain our 't~Hooft anomaly, partly because quark fields in the QCD Lagrangian have fractional charges under the baryon number symmetry $U(1)_\rmB$. 
By keeping the manifest one-form invariance in UV regularization, we can obtain the anomaly polynomial by applying the descent procedure to the Stora-Zumino chain. 

We discuss how the new anomaly matching condition is satisfied in the ordinary chiral-symmetry broken phase with quark bilinear condensate. 
Since our 't~Hooft anomaly involves $U(1)_\rmB$, it cannot be matched by the Wess-Zumino term of the pion Lagrangian. 
We find that the nontrivial topology of the vacuum manifold, $\pi_3(SU(N_\rmf))=\mathbb{Z}$, is important to match the anomaly, and this is nothing but the unified description of nucleons and mesons by Skyrme. 
Furthermore, conservation of the baryon number current $J_\rmB$ must be anomalously broken under the background gauge fields for $(\mathbb{Z}_{N_\rmf})_\rmL$ and $SU(N_\rmf)_\rmV/\mathbb{Z}_{N_\rmf}$ in order to match the new 't~Hooft anomaly. 
This is indeed satisfied for the ordinary chiral broken phase. 

We also examine the exotic chiral-symmetry broken phase proposed by Stern, and the naive Stern phase is ruled out due to the mismatch of our 't~Hooft anomaly. 
This phase is characterized by unbroken discrete chiral symmetry, and the vacuum manifold is $PSU(N_\rmf)=SU(N_\rmf)/\mathbb{Z}_{N_\rmf}$. 
Although one can construct the skyrmion current $J_\rmB$ because $\pi_3(PSU(N_\rmf))=\pi_3(SU(N_\rmf))=\mathbb{Z}$, anomalous violation of $J_\rmB$ does not take the appropriate form because of the absence of discrete chiral symmetry $(\mathbb{Z}_{N_\rmf})_\rmL$ in the phenomenological Lagrangian. 
A previous study rules out this phase by QCD inequalities, and it is consistent with our result. We would like to emphasize that our result applies to much wider regions of QCD phase diagrams such as finite-density zero-temperature QCD, because the argument relies only on symmetry and anomaly. 

Since many phases of finite-density QCD have been proposed~\cite{Alford:2001dt, Fukushima:2010bq}, it is useful to consider about the consistency with the anomaly matching condition in order to restrict possible phases. This problem has been discussed in the context of the original 't~Hooft anomaly matching~\cite{Sannino:2000kg}, and it is an interesting problem to apply the new discrete 't~Hooft anomaly. 
In a previous study~\cite{Tanizaki:2017mtm}, the color-flavor locked phase for $N_\rmc=N_\rmf=3$ is shown to satisfy the discrete anomaly matching by breaking the $(\mathbb{Z}_{2N_\rmf})_\rmA$ in the context of QCD with symmetry-twisted boundary conditions, and we can now consider the similar problem at generic numbers of $N_\rmc$ and $N_\rmf$ for zero-temperature finite-density QCD. 
This is an important future work. 

Lastly, we checked the anomaly matching between the dual descriptions of $\mathcal{N}=1$ SQCD with $N_\rmf\ge N_\rmc+1$. 
For $N_\rmf\ge N_\rmc+2$, Seiberg duality claims that $SU(N_\rmc)$ gauge theory is mapped to $SU(N_\rmf-N_\rmc)$ gauge theory, and explicit computation of the new 't~Hooft anomaly of both sides gives the same result.  
When $N_\rmf=N_\rmc+1$, the $s$-confining phase is realized, i.e. confinement occurs without chiral symmetry breaking. The massless excitations are color-singlet mesons and chiral baryons, and the new anomaly is matched by massless chiral baryons. 
We do not discuss the case $N_\rmf\le N_\rmc$ in this paper, so let us just give a brief comment: For $N_\rmf=N_\rmc$, we expect that the anomaly is matched as in the case of non-SUSY QCD discussed in this paper, when we choose the chiral-symmetry broken phase of quantum moduli. For $N_\rmf<N_\rmc$, SQCD has a runaway vacuum, and we are not sure whether the anomaly matching makes sense in such a situation.

In this paper, we only consider the 't~Hooft anomaly of internal symmetry but we can extend this argument using the spacetime symmetry by including the background gravity. 
For example, since the internal symmetry include $U(1)_\rmV$, we can consider the structure group as $[\mathrm{Spin}(4)\times U(1)_\rmV]/\mathbb{Z}_2$, where $\mathrm{Spin}(4)$ is the spacetime symmetry group. 
This is a $\mathrm{Spin}^c$ structure, and it allows us to put massless QCD on non-spin manifolds to detect discrete gauge-gravitational anomalies if they exist. Studying such anomaly matching gives a new constraint on possible QCD vacua, and it must be an interesting future study.


\acknowledgments
The author thanks Zohar Komargodski and Erich Poppitz for reading early drafts of this paper and giving useful comments.  
The author especially thanks Erich Poppitz for suggesting a possibility to apply anomaly matching to the Stern phase.  Y.~T. is financially supported by RIKEN special postdoctoral program. 


\appendix
\section{Alternative derivation of the anomaly linear in $A_\chi^{(1)}$}\label{app:alternative_derivation}

In this section, we give an alternative derivation of the linear term (\ref{eq:new_anomaly}) in $A_\chi^{(1)}$ of $S_{\mathrm{SPT}}$. 
Here, we only introduce the background gauge field $(A_\rmf, A_\rmV, B_\rmc^{(2)}, B^{(2)}_\rmf)$ of the vector-like symmetry $[SU(N_\rmf)_\rmV\times U(1)_\rmV]/[\mathbb{Z}_{N_\rmc}\times \mathbb{Z}_{N_\rmf}]$. 
After that, we will find that the discrete axial symmetry $(\mathbb{Z}_{2N_\rmf})_\rmA$, or $(\mathbb{Z}_{N_\rmf})_\rmL$, is broken by Abelian anomaly. 
This strategy is useful to find the mixed 't~Hooft anomaly, and used in some previous studies~\cite{Gaiotto:2017yup, Tanizaki:2017bam, Komargodski:2017smk, Shimizu:2017asf, Tanizaki:2017mtm} etc. 

Introducing the vector-like background gauge fields, we obtain the quark kinetic term as 
\be
\overline{\psi}\gamma_{\mu}D_\mu\psi \Rightarrow 
\overline{\psi}\gamma_{\mu}\left(\p_{\mu}+[\widetilde{a}+\widetilde{A}_\rmf+A_\rmV]_{\mu}\right)\psi. 
\ee
We then define the partition function $Z[A_\rmf,A_\rmV,B^{(2)}_\rmc,B^{(2)}_\rmf]$. Let us perform the discrete axial transformation, $\psi \mapsto \rme^{2\pi \im/(2N_\rmf)\gamma_5}\psi$ and $\overline{\psi}\mapsto \overline{\psi}\rme^{2\pi \im/(2N_\rmf)\gamma_5}$, then the action is invariant, but the path integral measure changes as 
\bea
\Diff \overline{\psi}\Diff \psi&\mapsto& 
\Diff \overline{\psi}\Diff \psi \exp\left({2\pi\im \over 2N_\rmf}{2\over 8\pi^2}\int \tr_{\rmc,\rmf}\left[(\widetilde{F}_\rmc+\widetilde{F}_\rmf+\diff A_\rmV)^2\right]\right)\nonumber\\
&=&\Diff \overline{\psi}\Diff \psi \exp \im\left({N_\rmc/N_\rmf\over 4\pi}\int  \tr_\rmf \left[\widetilde{F}_\rmf^2\right]+
{1\over 2\pi}\int \diff A_\rmB \wedge B^{(2)}_f\right). 
\eea
Since the extra phase is independent of the dynamical field, this is an extra phase of the partition function after performing the discrete axial symmetry. This 't~Hooft anomaly is nothing but the one given by (\ref{eq:new_anomaly}).

\bibliographystyle{utphys}
\bibliography{QFT}
\end{document}